\documentclass[sn-mathphys]{sn-jnl}

\newcommand{\gsim}{\, \raisebox{-0.8ex}{$\stackrel{\textstyle >}{\sim}$ }}

\usepackage{ulem}

\jyear{2021}%

\theoremstyle{thmstyleone}%
\theoremstyle{thmstyletwo}%

\theoremstyle{thmstylethree}%

\begin{document}

\title[Accuracy of 1-dimensional approximation in NS QNMs]{Accuracy of one-dimensional approximation in neutron star quasi-normal modes}


\author*[1,2]{\fnm{Hajime} \sur{Sotani}}\email{sotani@yukawa.kyoto-u.ac.jp}

\affil*[1]{\orgdiv{Astrophysical Big Bang Laboratory}, \orgname{RIKEN}, \orgaddress{\street{2-1 Hirosawa}, \city{Wako}, \postcode{351-0198}, \state{Saitama}, \country{Japan}}}

\affil[2]{\orgdiv{Interdisciplinary Theoretical \& Mathematical Science Program (iTHEMS)}, \orgname{RIKEN}, \orgaddress{\street{2-1 Hirosawa}, \city{Wako}, \postcode{351-0198}, \state{Saitama}, \country{Japan}}}


\abstract{
Since the eigenfrequency of gravitational waves from cold neutron stars becomes a complex number, where the real and imaginary parts respectively correspond to an oscillation frequency and damping rate, one has to somehow solve the eigenvalue problem concerning the eigenvalue in two-dimensional parameter space. To avoid this bother, one sometimes adopts an approximation, where the eigenvalue is in one-dimensional parameter space. In this study, first, we show the accuracy of the zero-damping approximation, which is one of the one-dimensional approximations, for the fundamental and 1st pressure modes. But, this approximation is not applicable to the spacetime mode, because the damping rate of the spacetime mode is generally comparable to the oscillation frequency. Nevertheless, we find the empirical relation for the ratio of the imaginary part to the real part of the eigenfrequency, which is expressed as a function of the steller compactness almost independently of the adopted equations of state for neutron star matter. Adopting this empirical relation, one can express the eigenfrequency in terms of just the real part, i.e., the problem to solve becomes an eigenvalue problem with a one-dimensional eigenvalue. Then, we find that the frequencies are estimated with good accuracy even with such approximations even for the 1st spacetime mode.
}




\maketitle

\section{Introduction}
\label{sec:I}

Neutron stars, which are produced through core-collapse supernovae, are one of the most suitable environments for probing physics under extreme conditions, which are quite difficult to realize on the Earth. The density inside the star significantly exceeds the standard nuclear density, while the gravitational and magnetic fields become much stronger than those observed in our solar system. So, by observing the neutron star itself and/or the phenomena associated with neutron stars, inversely one would be able to extract the information for the extreme conditions. For example, the discovery of the $2M_\odot$ neutron stars excludes some of the soft equations of state (EOSs) \cite{D10, A13, C20}. Namely, if the maximum mass of the neutron star constructed with an EOS does not reach the observed mass, such an EOS is not good for the EOS describing neutron stars. The light bending due to the strong gravitational field, which is one of the relativistic effects, may also tell us the neutron star properties. That is, owing to the strong gravitational field induced by the neutron star, the light curve from the neutron star would be modulated. Thus, by carefully observing the pulsar light curve, one could mainly constrain the stellar compactness, i.e., the ratio of the stellar mass to the radius (e.g., \cite{PFC83, LL95, PG03, PO14, SM18, Sotani20a}). In practice, the mass and radius of the neutron stars, i.e., PSR J0030+0451 \cite{Riley19, Miller19} and PSR J0740+6620 \cite{Riley21, Miller21}, are successfully constrained through the observations with the Neutron star Interior Composition Explorer (NICER) on the International Space Station. The current uncertainty in the mass and radius constraints is still large, but this type of constraint must help us to know the true EOS when the uncertainty is reduced through future observations.

In addition to the direct observations of the stellar mass and radius, the oscillation frequency from the neutron star is another important observable. Since the oscillation frequency strongly depends on the interior properties of the object, as an inverse problem, one can extract the stellar properties by observing the frequency. This technique is known as asteroseismology, which is similar to seismology on the Earth and helioseismology on the Sun. In fact, by identifying the quasi-periodic oscillations observed in the afterglow following the giant flares with the neutron star crustal oscillations, one could constrain the crust properties (e.g., Refs. \cite{GNHL2011, SNIO2012, SIO2016}). Similarly, one may know the neutron star mass, radius, and EOS by observing the gravitational wave frequencies from neutron stars and by identifying them with specific oscillation modes  (e.g., Refs. \cite{AK1996,AK1998,STM2001,SH2003,SYMT2011,PA2012,DGKK2013,Sotani20b,Sotani20c,Sotani21,SD22}). Furthermore, this technique is recently adopted for understanding the gravitational wave signals appearing in the numerical simulations for core-collapse supernova (e.g., Refs. \cite{FMP2003,FKAO2015,ST2016,SKTK2017,MRBV2018,SKTK2019,TCPOF19,SS2019,ST2020,STT2021}).

In general, the gravitational wave frequencies from the neutron stars are complex numbers, where the real and imaginary parts respectively correspond to an oscillation frequency and damping rate, and the corresponding eigenmodes are called quasi-normal modes. So, to determine the quasi-normal modes of the neutron stars, one has to somehow solve the eigenvalue problem with respect to the eigenvalue in two-dimensional parameter space with the real and imaginary parts. Since this solution may be a bother, one sometimes adopts the approximation to estimate the frequency of gravitational waves. The simple approximation is the relativistic Cowling approximation, where the metric perturbations are neglected during the fluid oscillations, i.e., the frequencies of fluid oscillations are determined with the fixed metric. One can qualitatively discuss the behavior of frequencies with the Cowling approximation, even though the accuracy of the determined frequencies is within $\sim 20\%$ \cite{ST2020, YK97}.  Another approximation adopted so far is the zero-damping approximation, where one takes into account the metric perturbations as well as the fluid perturbations, but the imaginary part of the eigenvalue is assumed to be zero. This is because the damping rate, i.e, the imaginary part of the eigenvalue, for the gravitational waves induced by the fluid oscillations is generally much smaller than the oscillation frequency, i.e., the real part of the eigenvalue. This approximation is considered to estimate the frequency of the gravitational waves well, but it is not discussed how well this approximation works. So, in this study, we first discuss how well the zero-damping approximation works by comparing the frequencies determined with the approximation to those determined through the proper eigenvalue problem. Anyway, with either the Cowling approximation or the zero-damping approximation, one can discuss only the frequency of the gravitational waves induced by the fluid oscillations, such as the fundamental, pressure, and gravity modes, but one cannot discuss the gravitational waves associated with the oscillations of the spacetime itself, i.e., the $w$-modes.

In order to estimate the $w_1$-mode frequency, in this study, we find the empirical relation for the ratio of the damping rate of the $w_1$-mode gravitational wave, i.e., the imaginary part of the quasi-normal mode, to its frequency, i.e., the corresponding real part, as a function of the stellar compactness almost independently of the adopted EOSs. With this empirical relation, we newly propose a one-dimensional approximation for estimating the $w_1$-mode frequency and show how well this approximation works. Unless otherwise mentioned, we adopt geometric units in the following, $c=G=1$, where $c$ and $G$ denote the speed of light and the gravitational constant, respectively, and the metric signature is $(-,+,+,+)$.

\section{Neutron star models}
\label{sec:EOS}

\begin{figure}[tbp]
\centering
\includegraphics[scale=0.5]{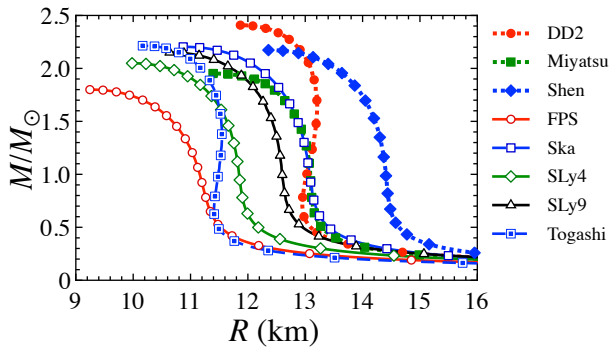}
\vspace{0.5cm}
\caption{
For the neutron star models constructed with various EOSs, the mass is shown as a function of the radius. Figure is taken from Ref. \cite{Sotani20c}.
}
\label{fig:MR}
\end{figure}

In this study, we simply consider the static, spherically symmetric stars, where the metric describing the system is given by 
\begin{equation} 
  ds^2 = -e^{2\Phi}dt^2 + e^{2\Lambda}dr^2 + r^2\left(d\theta^2 + \sin^2\theta d\phi^2\right).  \label{eq:metric}
\end{equation}
The metric functions $\Phi$ and $\Lambda$ in Eq. (\ref{eq:metric}) are functions only of $r$, while $e^{2\Lambda}$ is directly related to the mass function, $m(r)$, as $e^{-2\Lambda}=1-2m/r$. The stellar structure is determined by integrating the Tolman-Oppenheimer-Volkoff equation with an appropriate EOS for neutron star matter. In this study, we adopt the same EOSs as in Refs. \cite{Sotani20c, Sotani21}, which are listed in Table \ref{tab:EOS} together with the EOS parameters and maximum mass of the neutron star. Here, $K_0$ and $L$ are the incompressibility for symmetric nuclear matter and the density dependence of the nuclear symmetry energy. We note that any EOSs can be characterized by these nuclear saturation parameters, which are constrained via terrestrial nuclear experiments, such as $K_0= 230\pm 40$ MeV \cite{KM13} and $L\simeq 58.9\pm 16$ MeV \cite{Li19}. Comparing these fiducial values of $K_0$ and $L$ to those for the adopted EOSs, some of the EOSs listed in Table \ref{tab:EOS} seem to be excluded, but we consider even such EOSs in this study to examine the EOS dependence in the wide parameter space. Additionally, $\eta$ in Table \ref{tab:EOS} is the specific combination of $K_0$ and $L$ as $\eta\equiv \left(K_0 L^2\right)^{1/3}$ \cite{SIOO14}, which is a suitable parameter not only for expressing the properties of low-mass neutron stars but also for discussing the maximum mass of neutron stars \cite{SSB16,Sotani17,SK17}. The mass and radius relations for the neutron star models constructed with the EOS listed in Table \ref{tab:EOS} are shown in Fig. \ref{fig:MR}, where the filled and open marks respectively correspond to the EOSs constructed with the relativistic framework and with the Skyrme-type effective interaction, while the double-square is the EOS constructed with a variational method. As in Ref. \cite{Sotani20c}, we consider the stellar models denoted with marks in this figure.

\begin{table}
\caption{EOS parameters adopted in this study, $K_0$, $L$, and $\eta$, and the maximum mass, $M_{\rm max}$, of the neutron star constructed with each EOS.} 
\label{tab:EOS}
\begin {center}
\begin{tabular}{ccccc}
\hline\hline
EOS & $K_0$ (MeV) & $L$ (MeV) & $\eta$ (MeV) & $M_{\rm max}/M_\odot$   \\
\hline
DD2
 & 243 & 55.0  & 90.2  & 2.41  \\ 
Miyatsu
 & 274 &  77.1 & 118  & 1.95  \\
Shen
\ & 281 & 111  & 151  &  2.17  \\  
FPS
 & 261 & 34.9 & 68.2  & 1.80  \\  
SKa
 & 263 & 74.6 & 114 & 2.22  \\ 
SLy4
 & 230 & 45.9 &  78.5 & 2.05  \\ 
SLy9
 & 230 & 54.9 &  88.4 & 2.16  \\  
Togashi
 & 245  & 38.7  & 71.6 & 2.21  \\ 
\hline \hline
\end{tabular}
\end {center}
\end{table}

\section{Quasi-normal modes}
\label{sec:QNMs}

In order to determine the quasi-normal modes of gravitational waves from compact objects, one has to solve the eigenvalue problem. The perturbation equations are derived from the linearized Einstein equation by adding the metric and fluid perturbations on the background stellar models. By imposing the appropriate boundary conditions at the stellar center, surface, and spacial infinity, the problem to solve becomes the eigenvalue problem with respect to the eigenvalue, $\omega$. As a practical matter, how to numerically deal with the boundary condition at the spatial infinity may become a problem. In this study, we especially adopt the continued fraction method, the so-called Leaver's method \cite{Leaver85,Leaver90}. That is, the perturbation variable outside the star is expressed as a power series around the stellar surface, which also satisfies the boundary condition at the infinity. By substituting this expansion into the Regge-Wheeler equation, one can get a three-term recurrence relation including $\omega$, which is rewritten in the form of a continued fraction. So, one has to find the eigenvalue, $\omega$, which satisfies the continued fraction. Here, we symbolically express this condition as $f(\omega)=0$. The resulatant $\omega$ is generally a complex value, where the real and imaginary parts correspond to the oscillation frequency,  Re($\omega$)/$2\pi$, and damping rate of the corresponding gravitational waves. The concrete perturbation equations, boundary conditions, and functional form of $f(\omega)$ are shown in Refs. \cite{STM2001,ST2020}. 

In practice, since $f(\omega)$ is also a complex value, we try to find the value of $\omega$ numerically, with which the absolute value of $f(\omega)$ becomes the local minimum. In this study, we especially focus on the fundamental ($f$-), 1st pressure ($p_1$-), and 1st spacetime ($w_1$-) modes for the various neutron star models constructed with the EOSs listed in Table \ref{tab:EOS}. We note that the $f$- and $p_1$-modes are the quasi-normal modes induced by the fluid oscillations, while the $w_1$-mode is the quasi-normal mode associated with the oscillations of spacetime itself.

\begin{figure}[tbp]
\centering
\includegraphics[scale=0.5]{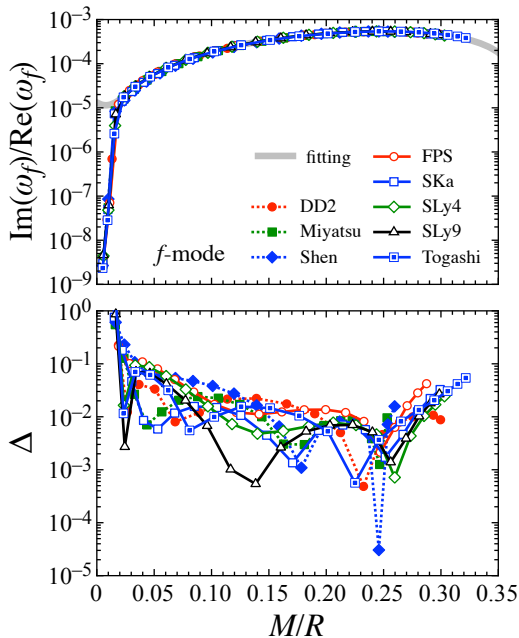}
\vspace{0.5cm}
\caption{
In the top panel, the ratio of Im($\omega$) to Re($\omega$) for the $f$-mode is shown as a function of the stellar compactness for various neutron star models, where the solid line denotes the fitting formula given by Eq. (\ref{eq:fitting_ff}). In the bottom pane, the relative deviation of the value of Im($\omega$)/Re($\omega$) estimated with the empirical formula from that calculated via the eigenvalue problem is shown. 
}
\label{fig:ratio-ff}
\end{figure}

\begin{figure}[tbp]
\centering
\includegraphics[scale=0.5]{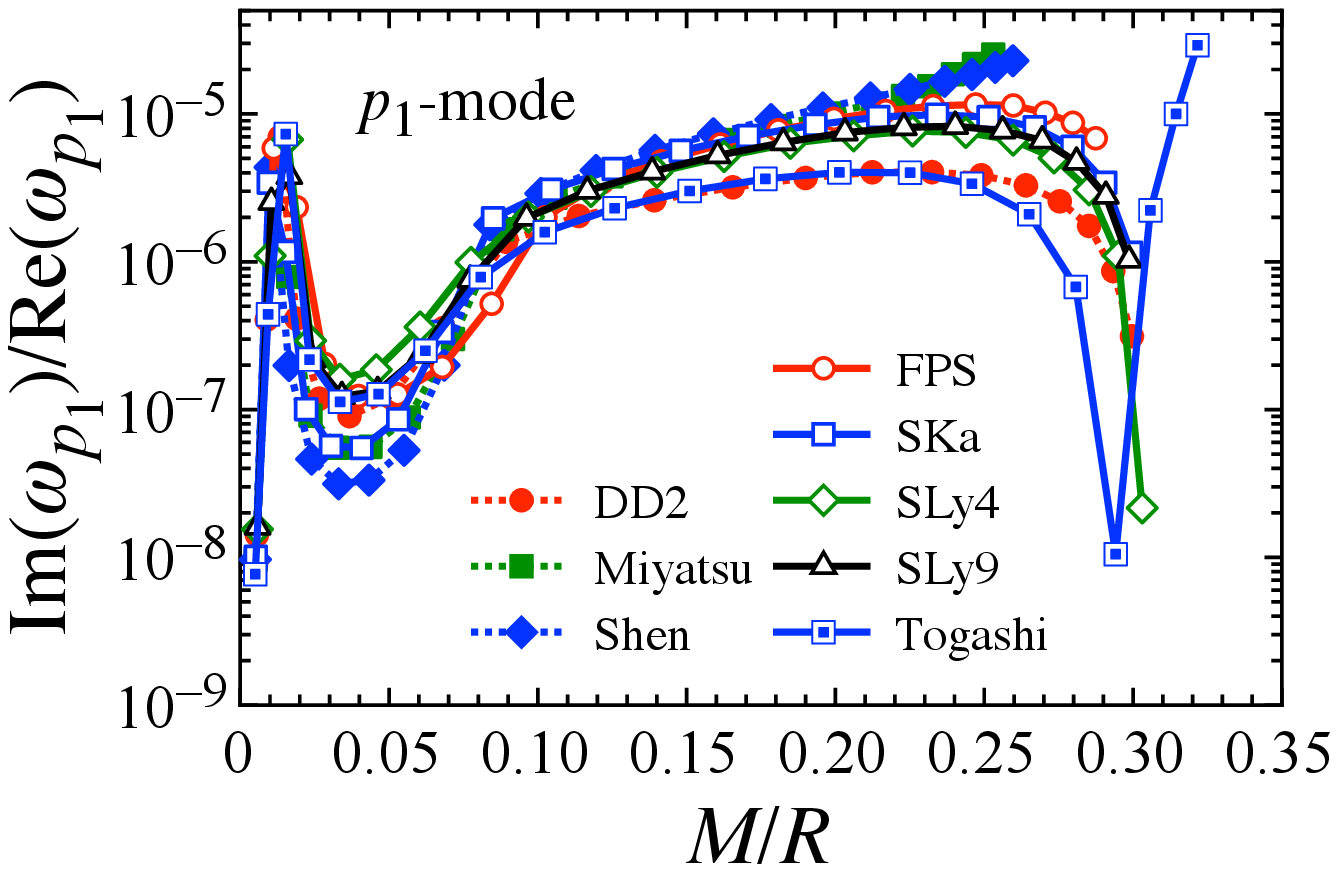}
\vspace{0.5cm}
\caption{
The ratio of Im($\omega$) to Re($\omega$) for the $p_1$-mode is shown as a function of the stellar compactness for various neutron star models.
}
\label{fig:ratio-p1}
\end{figure}

With the resultant $f$-mode frequency, we show the ratio of Im$(\omega)$ to Re$(\omega)$ in the top panel of Fig. \ref{fig:ratio-ff}. From this figure, one can observe that the values of Im$(\omega)$/Re$(\omega)$ are well expressed as a function of the stellar compactness, $M/R$, almost independently of the EOS. In fact, we can derive the empirical relation for Im$(\omega)$/Re$(\omega)\gsim 10^{-5}$, i.e., except for the quite low-mass neutron stars, given by 
\begin{equation}
  \frac{{\rm Im}(\omega_f)}{{\rm Re}(\omega_f)} = \left[0.13193 -4.4754\left(\frac{M}{R}\right)
        +290.9\left(\frac{M}{R}\right)^2 -756.14 \left(\frac{M}{R}\right)^3\right]\times10^{-4},\label{eq:fitting_ff}
\end{equation}
with which the expected values are shown with the thick-solid line in the top panel. In the bottom panel of Fig. \ref{fig:ratio-ff}, we also show the relative deviation calculated with
\begin{equation}
  \Delta = \frac{{\rm abs}[{\cal R}-{\cal R}_{\rm fit}]}{{\cal R}}, \label{eq:Delta}
\end{equation}
where ${\cal R}$ and ${\cal R_{\rm fit}}$ denote the ratio of Im$(\omega)$ to Re$(\omega)$ detemined through the eigenvalue problem and estimated with Eq. (\ref{eq:fitting_ff}) for each stellar model, respectively. From this figure, we find that the values of Im$(\omega)$/Re$(\omega)$ can be usually estimated within $\sim10\%$ accuracy by using the empirical relation. In a similar way, we show Im$(\omega)$/Re$(\omega)$ for the $p_1$-modes in Fig. \ref{fig:ratio-p1}. But, for the case of the $p_1$-mode we can not derive the empirical relation as a function of $M/R$, unlike the case of the $f$-mode.

\begin{figure}[tbp]
\centering
\includegraphics[scale=0.5]{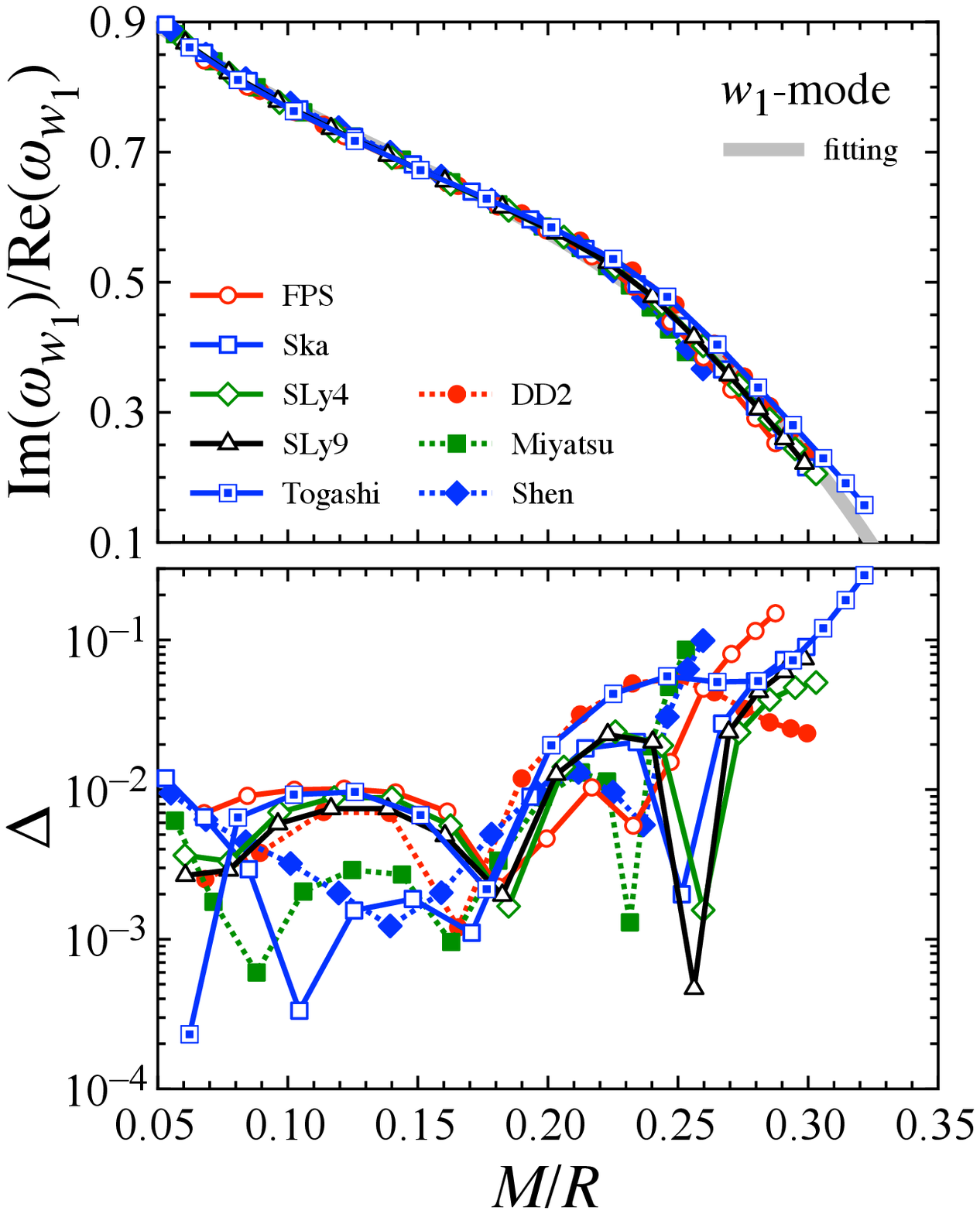}
\vspace{0.5cm}
\caption{
The ratio of Im$(\omega)$ to Re$(\omega)$ of the $w_1$-mode for various EOSs is shown as a function of the corresponding stellar compactness (top panel), where the solid line denotes the fitting line given by Eq. (\ref{eq:fitting_w1}). The bottom panel denotes the relative deviation of the estimation with the fitting formula from the values of Im$(\omega)$/Re$(\omega)$ calculated via the eigenvalue problem.
}
\label{fig:ratio}
\end{figure}

On the other hand, we also find that one can express the value of Im(${\omega}$)/Re($\omega$) for the $w_1$-mode as a function of $M/R$ almost independently of the adopted EOS. In the top panel of Fig. \ref{fig:ratio}, we show the ratio of Im$(\omega)$ to Re$(\omega)$ for the $w_1$-mode determined through the eivenvlue problem for each stellar model as a function of the corresponding stellar compactness, where the solid line denotes the fitting formula given by
\begin{equation}
  \frac{{\rm Im}(\omega_{w_1})}{{\rm Re}(\omega_{w_1})} = 1.0659 -4.1598\left(\frac{M}{R}\right)
        +16.4565\left(\frac{M}{R}\right)^2 -39.5369\left(\frac{M}{R}\right)^3.  \label{eq:fitting_w1}
\end{equation}
In the bottom panel, we show the relative deviation calculated with Eq. (\ref{eq:Delta}) for the $w_1$-mode, where ${\cal R}_{\rm fit}$ is estimated with Eq. (\ref{eq:fitting_w1}). Considering to the fact that $M/R=0.172$ (0.207) for the canonical neutron star model with $M=1.4M_\odot$ and $R=12$ km (10 km), one can observe that the fitting formula given by Eq. (\ref{eq:fitting_w1}) works better for the low-mass neturon star models.

\section{One-dimensional approximation}
\label{sec:Approximation0}

In order to determine the quasi-normal modes, one has to somehow search the solution of $\omega$ in two-dimensional parameter space with the real and imaginary parts, and this procedure may be trouble. One may sometimes adopt a suitable approximation to get out of this trouble, even if one can estimate only the frequency of quasi-normal modes. In this study, we especially consider the approximation, where the eigenvalue belongs to one-dimensional parameter space, depending only on the real part of the eigenvalue. We refer to this approximation as the one-dimensional approximation in this study. For example, since the imaginary part of the quasi-normal modes induced by the fluid oscillations is much smaller than the real part of them, as shown in Figs. \ref{fig:ratio-ff} and \ref{fig:ratio-p1}, one may assume that Im$(\omega)=0$. This approximation, referred to as zero-damping approximation, is a special case of one-dimensional approximation. In fact, this approximation has been adopted in some previous studies, but it has not been discussed how well this approximation works. So, in the next subsections, we will see the accuracy of the zero-damping approximation for the $f$- and $p_1$-mode frequencies, and then we also propose the one-dimensional approximation for estimating the $w_1$-mode frequency.

\subsection{Zero-damping approximation}
\label{sec:Approximation1}

\begin{figure}[tbp]
\centering
\includegraphics[scale=0.5]{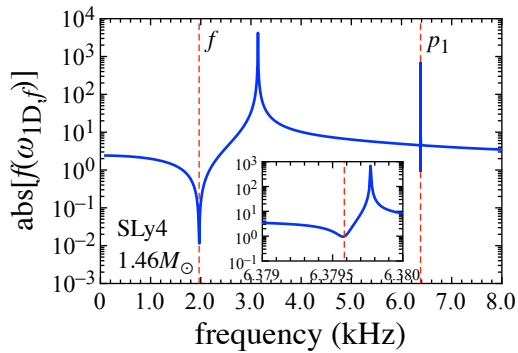}
\vspace{0.5cm}
\caption{
For the neutron star model with $1.46M_\odot$ constructed with SLy4 EOS, we show the absolute value of $f(\omega_{{\rm 1D},f})$ as a function of frequency. The vertical dashed lines denote the Re$(\omega)/2\pi$ for the $f$- and $p_1$-modes determined through the proper eigenvalue problem without the approximation. The panel inserted in the figure is an enlarged drawing in the vicinity of the $p_1$-mode frequency. 
}
\label{fig:SLy4-13}
\end{figure}

The zero-damping approximation, neglecting the imaginary part of the eigenvalue, is the simplest one-dimensional approximation, i.e., the eigenvalue, $\omega$, is assumed to be 
\begin{equation}
  \omega_{{\rm 1D},f} = \omega_{r}, \label{eq:omega_1D0}
\end{equation}
where $\omega_r$ is some real number and $\omega_{{\rm 1D},f}$ denotes the eigenvalue with the approximation for the gravitational wave induced by the fluid oscillations. With the zero-damping approximation, one can estimate the frequency, with which the absolute value of $f(\omega_{{\rm 1D},f})$ becomes the local minimum. Once the value of $\omega_r$ would be determined, the frequency of a gravitational wave is given by $\omega_r/2\pi$. As an example, we show the absolute value of $f(\omega_{{\rm 1D},f})$ as a function of the frequency for the neutron star model with $1.46M_\odot$ constructed with SLy4 EOS in Fig. \ref{fig:SLy4-13}, where the vertical dashed lines denote the frequencies of the $f$- and $p_1$-modes determined through the proper eigenvalue problem without the approximation and the inserted panel is an enlarged drawing in the vicinity of the $p_1$-mode frequency.

\begin{figure}[tbp]
\centering
\includegraphics[scale=0.5]{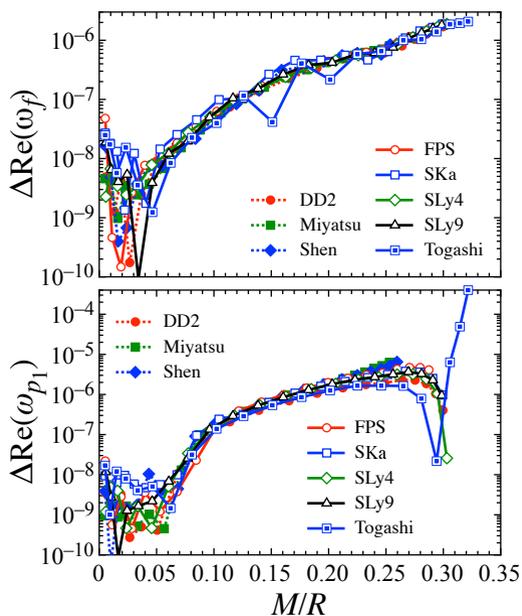}
\vspace{0.5cm}
\caption{
The relative deviation of the value of $\omega_r$ determined with the zero-damping approximation from the value of Re$(\omega)$ determined through the proper eigenvalue problem without the approximation, which is calculated with Eq. (\ref{eq:DRef}), is shown as a function of $M/R$ for various neutron star models. The top and bottom panels correspond to the results for the $f$- and $p_1$-modes, respectively. 
}
\label{fig:Dfp1}
\end{figure}

In order to check the accuracy of the frequencies of the $f$- and $p_1$-modes with the zero-damping approximation, we estimate the relative deviation of them from the corresponding frequencies determined through the proper eigenvalue problem, which is calculated by
\begin{equation}
  \Delta {\rm Re}(\omega_i) = \frac{{\rm abs}[{\rm Re}(\omega_i)-\omega_{r,i}]}{{\rm Re}(\omega_i)}, 
      \label{eq:DRef}
\end{equation}
where ${\rm Re}(\omega_i)$ and $\omega_{r,i}$ respectively denote the real part of $\omega$ determined through the proper eivenvalue problem without the approximation and the value of $\omega_r$ determined with the zero-damping approximation for the $f$- ($i=f$) and $p_1$-modes ($i=p_1$). For various neutron star models, we show the values of $\Delta {\rm Re}(\omega_i)$ for the $f$-mode ($p_1$-mode) in the top (bottom) panel of Fig. \ref{fig:Dfp1}. From this figure, one can observe that the zero-damping approximation works significantly well.

\subsection{One-dimensional approximation for the $w_1$-mode}
\label{sec:Approximation2}

\begin{figure}[tbp]
\centering
\includegraphics[scale=0.5]{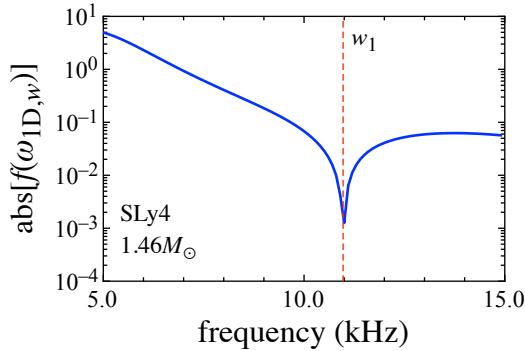}
\vspace{0.5cm}
\caption{
The absolute value of $f(\omega_{{\rm 1D},w})$ is shown as a function of the frequency for the neutron star model with $1.46M_\odot$ constructed with SLy4 EOS, where the vertical dashed line denotes the $w_1$-mode frequency determined though the proper eigenvalue problem without the approximation. 
}
\label{fig:SLy4-08}
\end{figure}

Unlike the gravitational waves induced by fluid oscillations, such as the $f$- and $p_1$-modes, the imaginary part of the spacetime modes ($w$-modes) becomes generally comparable to the real part of them, as shown in Fig. \ref{fig:ratio}. So, one cannot estimate the $w_1$-mode frequency with the zero-damping approximation. However, owing to the finding of the emprical relation for Im($\omega$)/Re($\omega$) as shown in Fig. \ref{fig:ratio}, we propose the one-dimensional approximation for the $w_1$-mode, i.e., the eigenvalue with the one-dimensional approximation is assumed to be
\begin{equation}
   \omega_{{\rm 1D},w}=\omega_r\left(1+{\rm i}{\cal R}_{\rm fit}\right), \label{eq:omega_1D1}
\end{equation}
where $\omega_r$ is some real value, while ${\cal R}_{\rm fit}$ denotes the ratio of Im$(\omega_{w_1})$ to Re$(\omega_{w_1})$ estimated with Eq. (\ref{eq:fitting_w1}). Then, one should find the suitable value of $\omega_r$, with which the absolute value of $f(\omega_{{\rm 1D},w})$ becomes local minimum. As an example, in Fig. \ref{fig:SLy4-08} we show the absolute value of $f(\omega_{{\rm 1D},w})$ as a function of the frequency given by $\omega_r/2\pi$ for the neutron star model with $1.46M_\odot$ constructed with SLy4 EOS. In this figure, for reference we also show the $w_1$-mode frequency determined through the proper eigenvalue problem with the dashed vartical line. 

\begin{figure}[tbp]
\begin{center}
\includegraphics[scale=0.5]{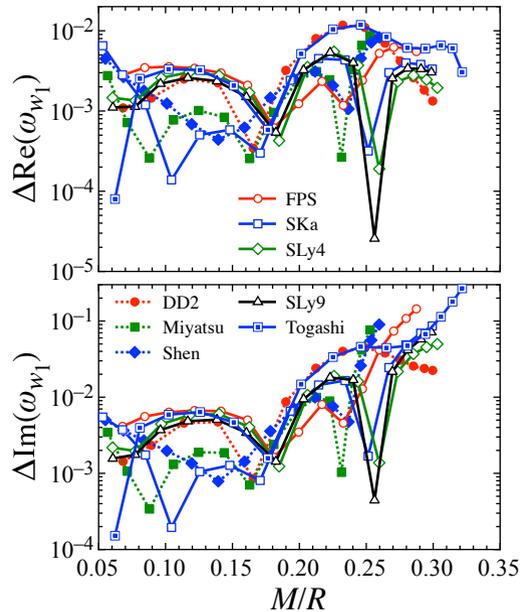}
\end{center}
\vspace{0.5cm}
\caption{
The relative deviation of the value of $\omega_r$ determined with the one-dimensional approximation from that of Re$(\omega)$ determined through the proper eigenvalue problem is shown as a function of stellar compactness for various neutron star models in the top panel. In the bottom panel, we show the relative deviation of the imaginary part of the eigenvalue estimated with the one-dimensional approximation from that of Im$(\omega)$ determined through the proper eigenvalue problem, calculated with Eq. (\ref{eq:DIm})
}
\label{fig:Dw1}
\end{figure}

To check the accuracy of the one-dimensional approximation, in the top panel of Fig. \ref{fig:Dw1} we show the relative deviation of the value of $\omega_r$ determined with the one-dimensional approximation from that of Re$(\omega)$ determined through the proper eigenvalue problem, calculated with Eq. (\ref{eq:DRef}) for the $w_1$-mode, as a function of $M/R$ for various neutron star models. In the bottom panel of Fig. \ref{fig:Dw1}, we also show the relative deviation, $\Delta {\rm Im}(\omega_{w_1})$, of the imaginary part of the eigenvalue estimated with the one-dimensional approximation, i.e., $\omega_r{\cal R}_{\rm fit}$, from Im$(\omega)$ determined through the proper eigenvalue problem, which is calculated with
\begin{equation}
  \Delta {\rm Im}(\omega_{w_1}) 
    = \frac{{\rm abs}[{\rm Im}(\omega_{w_1})- \omega_r{\cal R}_{\rm fit}]}{{\rm Im}(\omega_{w_1})}.
    \label{eq:DIm}
\end{equation}
From this figure, one can observe that the $w_1$-mode frequency can be estimated with the one-dimensional approximation within $\sim 1\%$ accuracy independently of the adopted EOSs. On the other hand, the damping rate of the $w_1$-mode can be estimated $\sim 30\%$ accuracy, which seems to strongly depend on the accuracy of the empirical relation for Im$(\omega_{w_1})$/Re$(\omega_{w_1})$ given by Eq. (\ref{eq:fitting_w1}).

\section{Conclusion}
\label{sec:Conclusion}

Quasi-normal modes are one of the important properties characterizing compact objects. In this study, first, we show that the ratio of the imaginary part to the real part of the quasi-normal mode for the $f$- and $w_1$-modes can be expressed as a function of the stellar compactness almost independently of the adopted EOSs, and we derive the corresponding empirical relations. Then, focusing on the $f$- and $p_1$-modes, which are gravitational wave frequencies induced by the stellar fluid oscillations, we examine the accuracy of the zero-damping approximation to estimate the corresponding frequencies of gravitational waves, where the damping rate, i.e., the imaginary part of the quasi-normal mode, is neglected. As a result, we show that one can estimate the frequencies of the $f$- and $p_1$-mode with the zero-damping approximation with considerable accuracy. In addition, we newly propose the one-dimensional approximation for estimating the $w_1$-mode frequency by adopting the empirical relation (that we find in this study) for the ratio of the imaginary part to the real part of the $w_1$-modes, and show that this approximation works well, where one can estimate the frequency within $\sim 1\%$ accuracy.

\bmhead{Acknowledgments}
This work is supported in part by the Japan Society for the Promotion of Science (JSPS) KAKENHI Grant Numbers 
JP19KK0354,  
JP20H04753,  and 
JP21H01088,  
and by Pioneering Program of RIKEN for Evolution of Matter in the Universe (r-EMU).

\section*{Declarations}

The author has no relevant financial or non-financial interests to disclose.

\end{document}